\documentclass[11pt]{article}
\usepackage{amsmath,amsfonts,latexsym,graphicx,amssymb}
\date{}

\def\r{\,\rangle}
\newcommand{\eeq}{\end{eqnarray}}
\newcommand{\beq}{\begin{eqnarray}}

\def\con{{}_{\_\rule{-1pt}{0pt}\_}
\rule{-2pt}{0pt}\raise1.5pt\hbox{$\mid$}\hspace{2pt}}

\def\oper{{\mathchoice{\rm 1\mskip-4mu l}{\rm 1\mskip-4mu l}%
{\rm 1\mskip-4.5mu l}{\rm 1\mskip-5mu l}}}

\title{Wigner function for damped systems}
\author{Dariusz Chru\'sci\'nski \\
 Institute of Physics, Nicholas Copernicus University\\
 ul. Grudzi\c{a}dzka 5/7, 87-100 Toru\'n, Poland\\
 e-mail: darch@phys.uni.torun.pl}

\begin{document}

\maketitle

\begin{abstract}
Both classical and quantum damped systems give rise to complex
spectra and corresponding resonant states. We investigate how
resonant states, which do not belong to the Hilbert space, fit the
phase space formulation of quantum mechanics. It turns out that
one may construct out of a pair of resonant states an analog of
the  stationary Wigner function.
\end{abstract}

\section{Introduction}
\setcounter{equation}{0}

Recently, it was observed by Kossakowski \cite{Kossak1},
\cite{Kossak} that quantized simple damped systems (e.g. damped
harmonic oscillator) give rise to discrete complex spectra. The
corresponding eigenvectors may be interpreted as resonant states.
Such states play important role in quantum mechanics and it is
widely believed that they are responsible for the irreversible
dynamics of physical systems (see e.g. \cite{Bohm}).
 In a recent paper \cite{Koopman} it was shown that the
 damping
behaviour in a classical system may be also interpreted as
appearance of resonant states for the corresponding Koopman
operator. Consider  a classical Hamiltonian system defined by the
Hamilton function $H : P \rightarrow \mathbb{R}$. Assuming that a
Hamiltonian flow is complete one introduces a self-adjoint
operator $L_{H}$ in the Hilbert space $L^2(P,d\mu)$, where $d\mu$
denotes the standard Liouville measure on $P$. This so called
Koopman operator is defined by
\begin{equation}\label{}
  L_H \, f := i\{f,H\}\ ,
\end{equation}
where $\{\ , \ \}$ denotes the Poisson bracket in the algebra of
classical observables $C^\infty(P)$. Studying simple examples of
classical damped systems we showed \cite{Koopman} that
corresponding Koopman operators have discrete complex spectra.
Clearly, the generalized eigenvectors do not belong to the Hilbert
space $L^2(P,d\mu)$ and the appropriate mathematical language to
deal with this situation is the rigged Hilbert space or the
Gelfand triplet (cf. e.g. \cite{RHS1},\cite{RHS2}), that is, any
generalized eigenvector belongs to a dual space $D^*$ where $D$ is
an appropriate space of test functions in $L^2(P,d\mu)$.

 In the present paper we shall
study the phase space formulation of quantum damped systems.
Clearly, this formulation, called also the deformation
quantization, is perfectly equivalent to the standard Hilbert
space approach. However, as we already mentioned, resonant states
lie outside the Hilbert space, and hence, it would be interesting
to find how they fit phase space approach. Any vector $\psi \in
{\cal H}$ gives rise to a Wigner function $W_\psi$ on a classical
phase space $P$.
 As was shown already by Wigner \cite{Wigner} this
function is real and produces marginal probability densities $\int
W_\psi(x,p)dx$ and  $\int W_\psi(x,p)dp$. The classical limit of
$W_\psi$ reproduces a classical probability distribution on $P$.
Moreover, if $\psi$ is an eigenvalue of the Hamilton operator
$\widehat{H}$, i.e. $\widehat{H}\psi=E\psi$, then the
corresponding Wigner function $W_\psi$ satisfies the following
eigenvalue problem:
\begin{equation}\label{}
  H \star W_\psi = W_\psi \star H = EW_\psi\ ,
\end{equation}
where $H$ is a classical Hamiltonian on $P$ and $f\star g$ denotes
``quantum deformation" of a usual commutative product of functions
$f\cdot g$ (see the next section).

It turns out that resonant states appear always in pairs:
\begin{equation}\label{}
  \widehat{H} \psi^\pm = E^\pm\psi^\pm\ ,
\end{equation}
such that
\begin{equation}\label{}
  E^- = \overline{E^+}\ .
\end{equation}
We show that each pair gives rise to a pair of function $F^\pm$ on
$P$ satisfying
\begin{equation}\label{}
H \star F^\pm = F^\pm \star H = E^\pm F^\pm\ ,
\end{equation}
with
\begin{equation}\label{}
  F^- = \overline{F^+}\ \ \   \   {\rm and} \ \ \ \ E^- = \overline{F^+}\ .
\end{equation}
Moreover, if $\psi^+=\psi^- =\psi\in {\cal H}$, then $E^+ =E^- \in
\mathbb{R}\,$ and  $\,F^+=F^-=W_\psi\,$, that is, $\,F^\pm\,$ may
be considered as a generalization of Wigner function for resonant
states. Functions $F^\pm$ do indeed satisfy basic properties of
Wigner function: they may be normalized so that $\int_P F^\pm\,
d\mu=1$, and they give rise to marginal probability distributions:
if $(x_1,\ldots,x_n,p_1,\ldots,p_n)$ are canonical coordinates on
$P$, then
\begin{eqnarray}\label{}
  \pi^\pm_{\bf x}(x_1,\ldots,x_n) := \int
  F^\pm(x_1,\ldots,x_n,p_1,\ldots,p_n)\, dp_1\ldots dp_n \ , \\
\pi^\pm_{\bf p}(p_1,\ldots,p_n) := \int
  F^\pm(x_1,\ldots,x_n,p_1,\ldots,p_n)\, dx_1\ldots dx_n \ ,
\end{eqnarray}
are classical probability distributions on $P$. Actually, in the
examples considered in this paper one finds
\begin{eqnarray}\label{}
 \pi^\pm_{\bf x}(x_1,\ldots,x_n) = \delta(x_1) \ldots \delta(x_n)\
 , \\
 \pi^\pm_{\bf p}(p_1,\ldots,p_n) = \delta(p_1) \ldots \delta(p_n)\
 .
\end{eqnarray}
It seems that the above result does contradict Heisenberg
uncertainty relations. Note, however, that resonant states do not
belong to the Hilbert space and hence the probabilistic
interpretation is not clear.

 In the next section we recall phase space approach to quantum
mechanics. Section~3 shows how this approach works for the
harmonic oscillator. Following sections discuss damped systems: a
toy model of a damped motion $\dot{x} = -\gamma x$, and damped
harmonic oscillator. We end up with some conclusions.

\section{Phase space formulation of quantum mechanics}
\setcounter{equation}{0}

Deformation quantization consists in replacing a commutative
algebra of functions $C^\infty(P)$ over a classical phase space
$P$ by a noncommutative algebra $(C^\infty(P),\star)$. For
simplicity let us assume that $P=\mathbb{R}^{2N}$. The
$\star$-product operation
\[  \star\ : \ C^\infty(P) \ \times \ C^\infty(P)\
\longrightarrow\ C^\infty(P) \]
 is defined by:
\begin{equation}\label{star}
  f\, \star \, g := f\,
  \exp\left[\frac{i\hbar}{2}\overleftrightarrow{\Lambda}
  \right]\, g  \ ,
\end{equation}
where $\Lambda$ denotes a bidifferential operator
\begin{equation}\label{}
  f\, \overleftrightarrow{\Lambda}\, g := \{f,g\}\ .
\end{equation}
The above structure was introduced long ago by Groenewold
\cite{Groenewold} and later on it was used by Moyal \cite{Moyal}
to construct the so called phase-space formulation of quantum
mechanics (see e.g. recent review by Zachos \cite{Zachos-r}). The
equivalence of the above approach to the standard Hilbert space
one is based on the well known Weyl correspondence: if
$\widehat{A}$ is self-adjoint operator on ${\cal H}=L^2(
\mathbb{R}^N,d{\bf x})$, then one defines the symbol $A({\bf
u},{\bf v})$ of the operator $\widehat{A}\,$ by:
\begin{equation}\label{Weyl1}
  A({\bf u},{\bf v}) := {\rm Tr} \left( \widehat{A} \, e^{(-i/\hbar) (
  {\bf u} \widehat{{\bf p}} + {\bf v}
\widehat{{\bf x}})} \right) \ ,
\end{equation}
where $\widehat{{\bf x}} = (\widehat{x}_1, \ldots ,
\widehat{x}_N)$ and $\widehat{{\bf p}}=(\widehat{p}_1, \ldots ,
\widehat{p}_N)$ are standard position and momentum operators in
$L^2(\mathbb{R}^N,d{\bf x})$. Conversely, given a symbol $A({\bf
u},{\bf v})$ one construct a corresponding operator:
\begin{equation}\label{Weyl2}
  \widehat{A} = \int d {\bf u} \int d{\bf v}
  \, A({\bf u},{\bf v})\,  e^{(-i/\hbar) (  {\bf u}\widehat{{\bf p}}
+ {\bf v} \widehat{{\bf x}})}\ .
\end{equation}
Now, if $\widehat{C} =\widehat{A} \widehat{B}$, then
\begin{equation}\label{}
  C = A \star B\ ,
\end{equation}
where $A,B,C$ are symbols of $\widehat{A}$, $\widehat{B}$ and
$\widehat{C}$, respectively. In this approach the von Neumann
equation for the density operator $\widehat{\rho}$
\begin{equation}\label{}
  i\hbar\, \partial_t \widehat{\rho} = [ \widehat{H},\widehat{\rho}]\ ,
\end{equation}
is replaced by the Moyal equation for the corresponding Wigner
function $W$, i.e. symbol of $\widehat{\rho}$:
\begin{equation}\label{Moyal}
   i\hbar\, \partial_t W = \{H,W\}_{\rm M} \ ,
\end{equation}
where the Moyal brackets is given by:
\begin{equation}\label{}
\{H,W\}_{\rm M} := H \star W - W \star H \ .
\end{equation}
 Using (\ref{Weyl1}) it is
easy to show that Wigner function corresponding to
$\widehat{\rho}$ is given by \cite{Wigner}
\begin{equation}\label{W}
  W({\bf x},{\bf p}) := \frac{1}{(2\pi)^N} \int d{\bf y}\, e^{-i{\bf py}}\,
  \Big\langle {\bf x} -
  \frac{\hbar}{2}{\bf y} \Big|\, \widehat{\rho}\, \Big| {\bf x} +
  \frac{\hbar}{2}{\bf y} \Big\rangle \ ,
\end{equation}
where $|{\bf x}\r$ is a  generalized eigenvector of $\widehat{{\bf
x}}$, i.e. $\widehat{{\bf x}}|{\bf x}\r = {\bf x}|{\bf x}\r$. This
way quantum mechanics may be formulated entirely in terms of
objects living on a classical phase space $P$. The very definition
of the $\star$-product implies
\begin{equation}\label{}
  f\star g - g \star f = i\hbar \{f,g\} + O(\hbar^2) \ ,
\end{equation}
and hence in the classical limit $\hbar \rightarrow 0$ the Moyal
equation (\ref{Moyal}) reproduces the Liouville equation
\begin{equation}\label{Lio}
  \partial_t f_{\rm cl} = \{ H,f_{\rm cl}\} \ ,
\end{equation}
for the classical density probability $f_{\rm cl}$ on $P$. Now,
the unitary evolution of $\widehat{\rho}$:
\begin{equation}\label{}
  \widehat{\rho}(t) = U(t)\, \widehat{\rho}\, U^{-1}(t)\ ,
\end{equation}
with $U(t)= \exp(- (i/\hbar)t\widehat{H})$, is replaced by the
following formula for $W(t)$:
\begin{equation}\label{UWU}
  W(t) = U_\star(t) \star W \star  U^{-1}_\star(t)\ ,
\end{equation}
where the so called $\star$--exponential $U_\star$ is defined by
\cite{Flato}:
\begin{eqnarray}\label{U-star}
  U_\star(t) &=& {\rm Exp}\left(- (it/\hbar){H}\right) \nonumber \\
  &:=&
  1 +  (-it/\hbar)H + \frac{1}{2!} (-it/\hbar)^2 H\star H +
  \frac{1}{3!} (-it/\hbar)^3 H\star H \star H + \ldots \ .
\end{eqnarray}
In the classical limit
\begin{equation}\label{}
  W(t)\ \longrightarrow\ f_{\rm cl}(t) = e^{-it L_H}\, f_{\rm cl}\ ,
\end{equation}
that is,
\begin{equation}\label{f-cl-time}
  f_{\rm cl}({\bf x},{\bf p},t) = f_{\rm cl}({\bf x}(-t),{\bf p}(-t),0)\ ,
\end{equation}
where ${\bf x}(t)$ and ${\bf p}(t)$ stand for the classical
evolution of $\bf x$ and $\bf p$, respectively. Actually, using
(\ref{star}), one finds that  the quantum evolution of $\bf x$ and
$\bf p$ (in the Heisenberg picture)
\begin{eqnarray}\label{}
\dot{\bf x} &=& \frac{{\bf x} \star H - H \star {\bf x}}{i\hbar} =
\{{\bf x},H\} \ , \\ \dot{\bf p} &=& \frac{{\bf p} \star H - H
\star {\bf p}}{i\hbar} = \{{\bf p},H\} \ ,
\end{eqnarray}
is the same as the classical one.

Finally, let us turn to  the energy spectrum. In the standard
approach one solves for the standard eigenvalue problem for the
quantum Hamiltonian $\widehat{H}$:
\begin{equation}\label{H-psi-E}
  \widehat{H}\psi = E \psi \ .
\end{equation}
It is easy to see that the corresponding Wigner function $W$
satisfies:
\begin{equation}\label{WHW}
  W_\psi \star H = H \star W_\psi = E W_\psi \ .
\end{equation}
Actually, one may prove (see e.g. \cite{Zachos}) that any real
solution $W$ of (\ref{WHW}) corresponds to a Wigner function for
$\psi$ satisfying (\ref{H-psi-E}). Moreover, if $\psi_n$ define an
orthonormal basis in $\cal H$, then corresponding Wigner functions
$W_n$ satisfy:
\begin{equation}\label{WnWm}
  W_n \star W_m = \frac{1}{(2\pi \hbar)^N}\ \delta_{nm} W_n \ ,
\end{equation}
and hence one obtains the following resolution of identity on $P$:
\begin{equation}\label{res}
  \sum_{n} W_n = \frac{1}{(2\pi \hbar)^N} \ ,
\end{equation}
which is phase space analog of the Hilbert space formula $
  \sum_{n} P_n = \oper$,
where $P_n$ is a 1-dimensional projector onto the eigenspace
generated by $\psi_n$. For more properties of Wigner function see
e.g. the review article \cite{Wigner-et-al}.

\section{Harmonic oscillator}
\label{OSCILLATOR}
 \setcounter{equation}{0}

To get the feeling how this approach works in practice let us
consider 1-dimensional harmonic oscillator described by the
following Hamiltonian:
\begin{equation}\label{H-ho}
  H_{\rm ho}(x,p) = \frac{\omega}{2} ( p^2 + x^2) \ .
\end{equation}
Now, let us study the corresponding eigenvalue problem
(\ref{WHW}). Equation $H \star W = EW$ gives:
\begin{equation}\label{osc1}
  \left[ x^2 + p^2 - \frac{\hbar^2}{4} ( \partial^2_x +
  \partial^2_p) + \frac{i\hbar}{2} (x\partial_p - p\partial_x) -
  \frac{2E}{\omega} \right] W = 0\ ,
\end{equation}
whereas $W \star H = EW$:
\begin{equation}\label{osc2}
  \left[ x^2 + p^2 - \frac{\hbar^2}{4} ( \partial^2_x +
  \partial^2_p) - \frac{i\hbar}{2} (x\partial_p - p\partial_x) -
  \frac{2E}{\omega} \right] W = 0\ .
\end{equation}
Therefore,
\begin{equation}\label{osc-zero}
(x\partial_p - p\partial_x)\, W = 0 \ ,
\end{equation}
which means that $W$ is a zero-mode of the Koopman operator
$L_{H_{\rm ho}}$ \cite{Koopman}. Taking into account (\ref{osc1})
and (\ref{osc2}) we obtain:
\begin{equation}\label{osc3}
  \left[ x^2 + p^2 - \frac{\hbar^2}{4} ( \partial^2_x +
  \partial^2_p)  -
  \frac{2E}{\omega} \right] W = 0\ .
\end{equation}
Introducing a new variable $\xi$:
\begin{equation}\label{}
  \xi := \frac{2}{\hbar}\, (x^2 + p^2)
\end{equation}
equation (\ref{osc3}) may be rewritten as follows:
\begin{equation}\label{osc4}
  \left[ \frac{\xi}{4} - \xi\partial^2_\xi - \partial_\xi -
  \frac{E}{\omega} \right] W(\xi) = 0 \ .
\end{equation}
Finally, defining $L=L(\xi)$ by
\begin{equation}\label{}
  W(\xi) =: e^{-\xi/2}\, L(\xi)\ ,
\end{equation}
equation (\ref{osc4}) implies:
\begin{equation}\label{LAG}
  \left[   \xi \partial^2_{\xi} + (1-\xi)\partial_\xi +
  \frac{E}{\hbar\omega} - \frac 12 \right] L(\xi) = 0 \ ,
\end{equation}
which is the defining equation of Laguerre's polynomials:
\begin{equation}\label{Ln}
  L_n(\xi) = \frac{1}{n!} \, e^\xi \partial_\xi(e^{-\xi}\, \xi)\ ,
\end{equation}
for $n= E/\hbar\omega - 1/2=0,1,\ldots\,$. This way one recovers
well known oscillator spectrum. The corresponding Wigner functions
$W_n$ read
\begin{equation}\label{Wn}
  W_n = \frac{(-1)^n}{\pi \hbar} \, e^{-\xi/2} \, L_n(\xi) \ .
\end{equation}
The reader will easily  check that $W_n$ defined in (\ref{Wn}) do
indeed satisfy formula (\ref{WnWm}).

It is well known that only $W_0$ which is given by the Gaussian
distribution
\begin{equation}\label{}
  W_0 = \frac{1}{\pi \hbar} \, e^{-\xi/2} = \frac{1}{\pi \hbar} \,
  e^{-(x^2+p^2)/\hbar}\ ,
\end{equation}
defines a probability distribution on $P$. However, in the
classical limit $\hbar \longrightarrow 0$ all  Wigner
 functions $W_n$ tend to well defined classical probability
 distributions. For example
\begin{equation}\label{W-limit}
  W_0(x,p) \ \longrightarrow \ \delta(x)\delta(p) \ .
\end{equation}

There is an alternative way to find the eigen-Wigner functions
$W_n$. One introduces phase-space analogs of creation and
annihilation operators:
\begin{equation}\label{}
  a = \frac{x+ip}{\sqrt{2\hbar}} \ , \ \ \ \ \ \ \
a^* = \frac{x-ip}{\sqrt{2\hbar}} \ ,
\end{equation}
satisfying  standard commutation relation:
\begin{equation}\label{}
\{a,a^*\}_{\rm M} =  a \star a^* - a^* \star a = 1 \ .
\end{equation}
It is easy to rewrite the formula for the $\star$-product
(\ref{star}) in terms of $a$ and $a^*$:
\begin{equation}\label{}
  f\star g = f\, e^{\frac 12 \left(
  \overleftarrow{\partial}\overrightarrow{\partial}^* -
  \overleftarrow{\partial}^*\overrightarrow{\partial}\right)}\, g
  \ ,
\end{equation}
where $\partial = \partial/\partial a$ and $\partial^*=
\partial/\partial a^*$. Oscillator Hamiltonian (\ref{H-ho}) takes
in the new variables the following form:
\begin{equation}\label{}
  H_{\rm ho} = \hbar \omega \left( a^* \star a + \frac 12 \right) \ .
\end{equation}
Now, let us define $W_0$ as a $\star$-Fock vacuum, that is,
\begin{equation}\label{Fock-vac}
  a \star W_0 = W_0 \star a^* = 0 \ ,
\end{equation}
and the corresponding excited states:
\begin{equation}\label{W-n-aa}
  W_n \ \propto\ a^{*n}  \star W_0 \star a^n  \ ,
\end{equation}
Noting that $\xi = 4|a|^2$ it is easy to check that $W_n$ defined
in (\ref{W-n-aa}) agrees with the formula  (\ref{Wn}).

Finally, let us turn to the time evolution defined in (\ref{UWU}).
The corresponding $\star$-exponential (\ref{U-star}) was found in
\cite{Flato} and is given by:
\begin{equation}\label{U-ho}
  {\rm Exp}(-(i/\hbar)tH_{\rm ho}) = \frac{1}{\cos(t/2)} \, \exp\left[ -2(i/\hbar)
  \tan(t/2) H_{\rm ho}\right] \ .
\end{equation}
Actually,  for the harmonic oscillator the quantum  evolution has
the same from as the classical one. It is evident from
(\ref{osc1}) and (\ref{osc2}) that
\begin{equation}\label{}
  \{H_{\rm ho},W\}_{\rm M} = i\hbar \{H_{\rm ho},W\} \ ,
\end{equation}
and hence, the Moyal equation (\ref{Moyal}) is the same as the
Liouville equation (\ref{Lio}). Therefore, due to
(\ref{f-cl-time})
\begin{equation}\label{W-osc-time}
  W({x},{p},t) = W({x}(-t),{p}(-t),0)\ .
\end{equation}

\section{Toy model of damped system}
\label{TOY} \setcounter{equation}{0}

Now, we apply this scheme to the simple damped system described by
the following equation:
\begin{equation}\label{damped}
  \dot{x} = - \gamma x \ ,
\end{equation}
where $\gamma >0$ is a damping constant. Clearly, this system is
not Hamiltonian. However, following \cite{Pontr} we may lift an
arbitrary dynamics on a configuration space $Q$
\begin{equation}\label{dotx-X}
  \dot{x} = {\bf X}(x)\ ,
\end{equation}
where $\bf X$ is a vector field on $Q$, to the Hamiltonian
dynamics on the corresponding phase space $P=T^*Q$. We define the
corresponding Hamiltonian
\begin{equation}\label{}
  H \ :\ P \ \longrightarrow\ \mathbb{R} \ ,
\end{equation}
by
\begin{equation}\label{}
  H(\alpha_x) := \alpha_x({\bf X}(x))\ ,
\end{equation}
for $\alpha_x \in T^*_xQ$. Using canonical coordinates
$(x_1,\ldots,x_N,p_1,\ldots,p_N)$ on $P$ we may rewrite a formula
for $H$ in a more familiar way:
\begin{equation}\label{}
  H(x,p) = \sum_{k=1}^N p_k \, X_k(x) \ .
\end{equation}
The corresponding Hamilton  equations read as follows:
\begin{eqnarray}\label{HAM-1}
  \dot{x}_k &=& \{ x_k,H\} = X_k(x) \ , \\
  \dot{p}_k &=& \{ p_k,H\} = - \sum_{l=1}^N p_l \frac{\partial
  X_l(x)}{\partial x_k} \ ,
\end{eqnarray}
for $k=1,\ldots,N$. In the above formulae $\{\ , \ \}$ denotes the
canonical Poisson bracket on $T^*Q$:
\begin{equation}\label{}
  \{F,G\} = \sum_{k=1}^N \left( \frac{\partial F}{\partial x_k}
   \frac{\partial G}{\partial p_k} - \frac{\partial G}{\partial x_k}
   \frac{\partial F}{\partial p_k}  \right) \ .
\end{equation}
Clearly, the formulae (\ref{HAM-1}) reproduce our initial
dynamical system (\ref{dotx-X}) on $Q$.

Now, applying the above procedure to (\ref{damped}) one obtains
the Hamiltonian system on $ \mathbb{R}^2$ with the Hamiltonian
given by:
\begin{equation}\label{H-damped}
  H_{\rm d}(x,p) = -\gamma xp\ .
\end{equation}
This system was analyzed in \cite{Koopman} where both {\em
classical spectrum} of the corresponding Koopman operator
$L_{H_{\rm d}}$ and quantum spectrum of
\begin{equation}\label{}
  \widehat{H}_{\rm d} = -\frac{\gamma}{2} \left(  \widehat{x}\widehat{p} +
  \widehat{p}\widehat{x} \right) \ ,
\end{equation}
were found:
\begin{equation}\label{}
  {\rm Spec}(L_{H_{\rm d}}) = \left\{ \, i\gamma n\ |\ n \in \mathbb{Z}\,
  \right\}\ ,
\end{equation}
and
\begin{equation}\label{}
  {\rm Spec}(\widehat{H}_{\rm d}) =
  \left\{ \, i\hbar\gamma \left( n+ \frac 12 \right) \ \Big|\
  n \in \mathbb{Z}\, \right\}\ .
\end{equation}
Both spectra are discrete and purely imaginary. It should be
stressed that both $L_{H_{\rm d}}$ and $\widehat{H}_{\rm d}$ are
self-adjoint operators on the corresponding Hilbert spaces $L^2(
\mathbb{R}^2)$ and $L^2( \mathbb{R})$, respectively. The
corresponding eigenvectors, which obviously  do not belong to the
Hilbert space, are usually called resonant states (see e.g.
\cite{Bohm}). It was found in \cite{Koopman}, \cite{Kossak} that
for
\begin{equation}\label{Psi-pm}
  \psi_n^+(x) : = x^n \ \ \ \ {\rm and} \ \ \ \ \psi_n^-(x):=
  (-i\hbar)^n\delta^{(n)}(x) \ , \ \ n=0,1,2,\ldots \ ,
\end{equation}
one has:
\begin{equation}\label{}
\widehat{H}\psi^{\pm}_n = \pm i\hbar\gamma \left( n + \frac 12
\right)  \psi^\pm_n \ .
\end{equation}
Evidently, these states living outside the Hilbert space $L^2(
\mathbb{R})$ can not be used to construct stationary Wigner
functions. Indeed, defining
\begin{equation}\label{}
  W^\pm_n(x,p) \ \propto\  \int dy e^{-ipy} \overline{\psi^\pm_n}(x -
  \frac{\hbar}{2} y) \, {\psi}^\pm_n(x +  \frac{\hbar}{2} y) \ ,
\end{equation}
one obtains
\begin{equation}\label{}
  W^\pm_n(x,p,t) = e^{\pm(2n+1)\gamma t}\, W^\pm_n(x,p,0) \ ,
\end{equation}
which shows that $W^\pm_n$ are non-stationary. To find the analogs
of stationary Wigner functions consider  the eigenvalue problem:
\begin{equation}\label{}
  H_{\rm d} \star F = EF  \ \ \ \ \ {\rm and} \ \ \ \ \ F\star H_{\rm d} = EF \
  .
\end{equation}
Equation $H_{\rm d} \star F = EF$ gives:
\begin{equation}\label{damped1}
  \left[ xp + \frac{\hbar^2}{4} \partial^2_{xp} +
  \frac{i\hbar}{2} (p\partial_p - x\partial_x) -
  \frac{E}{\gamma} \right] F = 0\ ,
\end{equation}
whereas $F \star H_{\rm d} = EF$:
\begin{equation}\label{damped2}
  \left[ xp + \frac{\hbar^2}{4}  \partial^2_{px}
 - \frac{i\hbar}{2} (p\partial_p - x\partial_x) -
  \frac{E}{\gamma} \right] F = 0\ .
\end{equation}
Therefore, in analogy to (\ref{osc-zero}) one finds
\begin{equation}\label{damped-zero}
(p\partial_p - x\partial_x)\, F = 0 \ ,
\end{equation}
which means that $F$ is a zero-mode of the corresponding Koopman
operator $L_{H_{\rm d}}$. Taking into account (\ref{damped1}) and
(\ref{damped2}) we obtain:
\begin{equation}\label{damped3}
  \left[ xp + \frac{\hbar^2}{4} \partial^2_{xp} +
  \frac{E}{\gamma} \right] F = 0\ .
\end{equation}
Introducing a new variable:
\begin{equation}\label{eta}
  \eta := \frac{4xp}{i\hbar}\ ,
\end{equation}
one may rewrite (\ref{damped3}) as follows:
\begin{equation}\label{}
  \left( \frac{\eta}{4} - \eta \partial^2_\eta - \partial_\eta -
  \frac{iE}{\hbar\gamma} \right) F(\eta) = 0 \ .
\end{equation}
Finally, defining $L(\eta)$:
\begin{equation}\label{}
  F(\eta) = e^{-\eta/2}L(\eta) \ ,
\end{equation}
one finds:
\begin{equation}\label{damped4}
  \left[ \eta\partial^2_\eta + (1-\eta) \partial_\eta - \left(
  \frac{iE}{\hbar\gamma} - \frac 12\right) \right] L(\eta) = 0 \ ,
\end{equation}
which is defining equation for  Laguerre polynomials (cf.
(\ref{LAG})). Hence we may define
\begin{equation}\label{Fn-}
  F^+_n = C_n \, e^{-\eta/2} L_n(\eta) \ , \ \ \ \ \ \
  n=0,1,2,\ldots\ ,
\end{equation}
where $n$th polynomial $L_n$ is given by (\ref{Ln}) and $C_n$ is a
normalization constant. The above formula for $F^+_n$ is an analog
of (\ref{Wn}) for the oscillator Wigner functions. Moreover, it
follows from (\ref{damped4}) that the spectrum is given by:
\begin{equation}\label{En}
  E_n = i\hbar\gamma \left( n + \frac 12 \right) \ , \ \ \ \ \ n= 0,1,2,\ldots\
  .
\end{equation}
Now, it is easy to check that functions $F^-_n$ defined by:
\begin{equation}\label{}
F^-_n :=  \overline{F^+_n}  \ ,
\end{equation}
satisfies
\begin{equation}\label{}
  H_{\rm d} \star F^-_n = F^-_n \star H_{\rm d} = - E_n F^-_n \ .
\end{equation}
It follows immediately from the following  property:
\begin{equation}\label{}
  \overline{f \star g} = \overline{g} \star \overline{f}\ ,
\end{equation}
which may be easily proved using the definition of the
$\star$-product (\ref{star}).

Now, let us study the basic properties of eigen-functions
$F^\pm_n$ and compare these with those of oscillator Wigner
functions $W_n$. Clearly, $F^\pm_n$ contrary to $W_n$ are not
real. Observe, that  taking a constant $C_n$ in (\ref{Fn-})
according to
\begin{equation}\label{Cn}
  C_n = \frac{(-1)^n}{\pi\hbar} \ ,
\end{equation}
i.e. like in (\ref{Wn}), one may prove that
\begin{equation}\label{}
  \int F^+_n(x,p)\,dxdp = \int F^-_n(x,p)\,dxdp = 1\ ,
\end{equation}
in perfect analogy to $W_n$. Moreover, $F^\pm_n$ give rise to the
following  marginal  probability distributions:
\begin{eqnarray}
  \int F^\pm_n(x,p)\,dx &=& \delta(p) \ , \\
 \int F^\pm_n(x,p)\,dp &=& \delta(x) \ .
\end{eqnarray}
This property seems to violate the Heisenberg uncertainty
principle -- the particle is localized both in $x$ and $p$
variables. Clearly, we lose the probabilistic interpretation of
$F^\pm_n$ since the corresponding eigenvectors $\psi^\pm_n$
(\ref{Psi-pm}) do not belong to the Hilbert space $L^2(
\mathbb{R})$. Interestingly, $F^\pm_n$ satisfy the following
condition:
\begin{equation}\label{FnFm}
  F^\pm_n \star F^\pm_m = \frac{1}{2\pi\hbar}\delta_{nm}\ F^\pm_n\ ,
\end{equation}
in perfect analogy to (\ref{WnWm}). Therefore, one obtains the
corresponding resolution of identity
\begin{equation}\label{}
  \sum_n F^\pm_n = 2\sum_n {\rm Re}\, F^+_n = \frac{1}{2\pi\hbar}\ .
\end{equation}

Finally, it would be interesting to find relation between the
resonant states $\psi^\pm_n$ defined in (\ref{Psi-pm}) and
$F^\pm_n$. Using some simple algebraic manipulations it is easy to
show that
\begin{equation}\label{}
  F^+_n(x,p) = C_n \int dy\, e^{-ipy} \overline{\psi^+_n}(x - \frac{\hbar}{2} y)\, \psi^-_n(x
  + \frac{\hbar}{2} y) \ ,
\end{equation}
and
\begin{eqnarray}\label{}
  F^-_n(x,p) &=& \overline{F^+_n}(x,p) = C_n \int dy\, e^{-ipy}
  \psi^+_n(x - \frac{\hbar}{2} y)\, \overline{\psi^-_n}(x + \frac{\hbar}{2} y) \nonumber \\
  &=& C_n \int dy\, e^{-ipy} \overline{\psi^-_n}(x - \frac{\hbar}{2} y)\, \psi^+_n(x
  + \frac{\hbar}{2} y) \ ,
\end{eqnarray}
with $C_n$ defined in (\ref{Cn}). Hence, each $F^\pm_n$ is built
out of $\psi^+_n$ and $\psi^-_n$. Note, that these eigenvectors
correspond to $E_n$ and $\overline{E}_n$, respectively. Clearly,
if $\psi$ is a proper eigenvectors corresponding to a real
eigenvalue $E$, then using the above prescription for $F$ one
recovers the Wigner function corresponding to $\psi$. Resonant
states comes always in pairs and two members of each pair are
needed to construct $F$.

Let us observe, that stationary functions $F^\pm_n$ may be defined
in a more transparent way. Rewriting the   Hamiltonian
(\ref{H-damped}) as follows:
\begin{equation}\label{}
  H_{\rm d}(x,p) = - \frac{\gamma}{2} ( x \star p + p \star x) \ ,
\end{equation}
let us define $F^+_0$ to be a normalized function satisfying the
following conditions:
\begin{equation}\label{}
  p \star F^+_0 = 0 \ \ \ \ \ {\rm and} \ \ \ \ \ F^+_0 \star x =
  0 \ ,
\end{equation}
which are solved by:
\begin{equation}\label{}
  F^+_0(x,p) = \frac{1}{\pi\hbar} \ e^{-2ixp/\hbar} \ .
\end{equation}
Having the ``$+$ ground state" $F^+_0$ one defines ``+ excited
states" by:
\begin{equation}\label{}
  F^+_n(x,p) \ \propto\ x^n \star F^+_0(x,p) \star p^n \ .
\end{equation}
Analogously, let us define $F^-_0$ to be a ``$-$ ground state"
satisfying:
\begin{equation}\label{}
 x \star F^-_0 = 0 \ \ \ \ \ {\rm and} \ \ \ \ \ F^-_0 \star p =
  0 \ .
\end{equation}
One finds
\begin{equation}\label{}
  F^-_0(x,p) = \frac{1}{\pi\hbar} \ e^{2ixp/\hbar} = \overline{F^+_0}(x,p)\ .
\end{equation}
The corresponding ``$-$ excited states" read:
\begin{equation}\label{}
  F^-_n(x,p) \ \propto\ p^n \star F^-_0(x,p) \star x^n \ .
\end{equation}
Using canonical commutation relation
\begin{equation}\label{}
  x \star p - p \star x = i\hbar\ ,
\end{equation}
one easily finds that $F^\pm_n$ do satisfy:
\begin{equation}\label{}
  H_{\rm d} \star F^\pm_n = F^\pm_n \star H_{\rm d} = \pm i\hbar\gamma \left(n +
  \frac 12 \right) \ .
\end{equation}

\section{Harmonic oscillator vs. damped system}
\setcounter{equation}{0}

Comparing the spectra of harmonic oscillator and damped system
considered in the previous section one finds striking similarity,
that is, they are related by the following relation:
\begin{equation}\label{}
  \omega = \pm i\gamma\ .
\end{equation}
Note, that performing the following canonical transformation:
\begin{equation}\label{}
  x =  \frac{1}{\sqrt{2}} (X+  P) \ \ \ \ \ {\rm
  and} \ \ \ \ \ p= \frac{1}{\sqrt{2}} (X -  P) \ ,
\end{equation}
one obtains
\begin{equation}\label{H-XP}
  H_{\rm d} = - \gamma xp = \frac{\gamma}{2} ( P^2 -  X^2) \ ,
\end{equation}
i.e. in the new variables $(X,P)$, $H_{\rm d}$ corresponds
formally to the harmonic oscillator with  $\omega = \pm i\gamma$.
This correspondence may be easily seen  by observing that both
Hamiltonians, i.e. $(P^2 + X^2)$ and $(P^2-X^2)$ are related by
the following $\star$-exponential:
\begin{equation}\label{}
  {V}_\lambda := {\rm Exp}(\lambda\, {XP}/\hbar) = 1 + \frac{\lambda}{\hbar}\, XP +
  \frac{\lambda^2}{2\hbar^2} \, XP \star XP + \ldots \ ,
\end{equation}
with $\lambda \in \mathbb{R}$. Indeed, one may show that
\begin{equation}\label{}
  V_{\lambda} \star X \star V_{-\lambda} = e^{-i\lambda}X \ \ \ \ \
  {\rm and } \ \ \ \ \
V_{\lambda} \star P \star V_{-\lambda} = e^{i\lambda}P \ .
\end{equation}
The above formulae imply:
\begin{equation}\label{}
 V_{\lambda} \star ( P^2 - X^2) \star V_{-\lambda} =
 e^{2i\lambda} (P^2 - e^{-4i\lambda}X^2) \ ,
\end{equation}
and hence, for $\lambda=\pm \pi/4$, one obtains:
\begin{equation}\label{}
 V_{\pm \pi/4} \star \left[ \frac{\gamma}{2}( P^2 - X^2)\right] \star V_{\mp \pi/4} =
 \pm \frac{i\gamma}{2} (P^2 + X^2) \ ,
\end{equation}
i.e. both systems are related by a {\em complex scaling} $V_{\pm
\pi/4}$. Therefore, it should be clear that the corresponding
eigen-functions $W_n$ and $F^\pm_n$ are also related by $V_{\pm
\pi/4}$. Let us denote
\begin{equation}\label{}
  H^\pm_{\rm ho} = \pm \frac{i\gamma}{2} (P^2 + X^2)\ .
\end{equation}
Now, if $W_n$ is an oscillator  Wigner function satisfying:
\begin{equation}\label{}
H^\pm_{\rm ho} \star W_n = W_n \star   H^\pm_{\rm ho} = \pm E_n
W_n \ ,
\end{equation}
with $E_n$ given by (\ref{En}), then $F^\pm_n$ defined by:
\begin{equation}\label{F-W}
  F^\pm_n := V_{\mp \pi/4} \star W_n \star V_{\pm \pi/4}\ ,
\end{equation}
satisfy the corresponding eigen-problem for the damped system:
\begin{equation}\label{}
H_{\rm d} \star F^\pm_n  = F^\pm_n \star H_{\rm d}  = \pm E_n
F^\pm_n\ .
\end{equation}
Moreover, it follows from (\ref{F-W}) that
\begin{equation}\label{}
  F^-_n = \overline{F^+_n} \ ,
\end{equation}
provided $W_n$ is real, and
\begin{equation}\label{}
  \int F^\pm_n \, dxdp = \int ( V_{\pm \pi/4} \star V_{\mp \pi/4})
  W_n\, dxdp = \int W_n \, dxdp\ ,
\end{equation}
i.e. $F^\pm_n$ are normalized on $ \mathbb{R}^2$.

Finally, let us observe that introducing on $ \mathbb{R}^2$ polar
coordinates $(r,\varphi)$ the corresponding Koopman operator
reads:
\begin{equation}\label{}
  L_{H_{\rm ho}} = i\omega\partial_\varphi \ ,
\end{equation}
and hence, oscillator Wigner functions are $SO(2)$ invariant since
$L_{H_{\rm ho}}W_n=0$. On the other hand using hyperbolic
coordinates $(s,\chi); \ P=s\cosh\chi, \ X=s\sinh\chi$, we obtain
\begin{equation}\label{}
  L_{H_{\rm d}} = i\gamma\partial_\chi\ ,
\end{equation}
and hence $F^\pm_n$ are $SO(1,1)$ invariant. Moreover, this
observation implies that the corresponding $\star$-exponential
$U_*(t)$ has for the damped system following form:
\begin{equation}\label{U-d}
  {\rm Exp}(-(i/\hbar)tH_{\rm d}) = \frac{1}{\cosh(t/2)} \, \exp\left[ -2(i/\hbar)
  \tanh(t/2) H_{\rm d}\right] \ ,
\end{equation}
which follows from (\ref{U-ho}). Note that
\begin{equation}\label{}
  \{H_{\rm d},F\}_{\rm M} = i\hbar\, \{H_{\rm d},F\}\ ,
\end{equation}
and hence, like for the harmonic oscillator, quantum and classical
evolution are given by the same formulae.

\section{Damped harmonic oscillator}
\label{DAMPED-HO} \setcounter{equation}{0}

Consider now a  damped harmonic oscillator described by the
following equation of motion:
\begin{equation}\label{}
  \ddot{x} + 2\gamma \dot{x} + \kappa x = 0 \ .
\end{equation}
 As is well known this
system plays a prominent role in various branches of physics,
especially in quantum optics. The above 2nd order equation may be
rewritten as a dynamical system on $ \mathbb{R}^2$
\begin{eqnarray}\label{}
  \dot{x}_1 &=& - \gamma x_1 + \omega x_2 \ , \\
  \dot{x}_2 &=& - \gamma x_2 - \omega x_1\ ,
\end{eqnarray}
with $\omega = \sqrt{\kappa - \gamma^2}$. Clearly this system is
not Hamiltonian if $\gamma \neq 0$. However, applying the
procedure of \cite{Pontr} one arrives at the following Hamiltonian
system on $ \mathbb{R}^4$:
\begin{eqnarray}\label{}
\dot{x}_1 &=& \{ x_1,H\} = -\gamma x_1 + \omega x_2 \ , \\
\dot{x}_2 &=& \{ x_2,H\} = -\omega x_1 - \gamma x_2 \ , \\
\dot{p}_1 &=& \{ p_1,H\} = +\gamma p_1 + \omega p_2 \ , \\
\dot{p}_2 &=& \{ p_1,H\} = -\omega p_1 + \gamma p_2 \ ,
\end{eqnarray}
where the corresponding Hamiltonian function is given by:
\begin{equation}\label{H-osc-d}
  H_{\rm dho}(x,p) = \omega( p_1x_2 - p_2x_1) - \gamma(p_1x_1 + p_2x_2)\ .
\end{equation}
Let us observe that the above Hamiltonian may be rewritten as
follows:
\begin{equation}\label{H-osc-d-1}
  H_{\rm dho}(x,p) = \omega( p_1\star x_2 - p_2\star x_1) -
  \frac{\gamma}{2}(p_1\star x_1 + x_1 \star p_1 + p_2\star x_2 + x_2 \star p_2)\ .
\end{equation}
Now, let us introduce a new set of variables:
\begin{eqnarray}\label{a1}
  a_1 &=& \frac{x_1 + ix_2}{\sqrt{2\hbar}} \ , \hspace{1.5cm}
a^*_1 \,=\, \frac{x_1 - ix_2}{\sqrt{2\hbar}} \ , \\   \label{a2}
  a_2 &=& \frac{ip_1 - p_2}{\sqrt{2\hbar}} \ , \hspace{1.5cm}
a^*_2 \,=\, \frac{-ip_1 - p_2}{\sqrt{2\hbar}} \ ,
\end{eqnarray}
satisfying the following commutation relations:
\begin{equation}\label{CCR1}
  \{ a_1,a_2\}_{\rm M} = \{ a_1,a^*_1\}_{\rm M} = \{
  a_2,a^*_2\}_{\rm M} =0 \ ,
\end{equation}
\begin{equation}\label{CCR2}
  \{ a_1,a^*_2\}_{\rm M} = \{ a_2,a^*_1\}_{\rm M} = 1\ .
\end{equation}
Hamiltonian (\ref{H-osc-d-1}) takes in new variables the following
form:
\begin{eqnarray}\label{H-d-osc-2}
H_{\rm dho} &=& \hbar \left( \alpha\, a^*_2\star a_1 +
\overline{\alpha}\,a^*_1\star a_2 + \omega \right) \nonumber\\ &=&
\hbar\alpha \left( a^*_2 \star a_1 + \frac 12 \right) +
\hbar\overline{\alpha} \left( a^*_1 \star a_2 + \frac 12 \right) \
,
\end{eqnarray}
where
\begin{equation}\label{}
  \alpha = \omega - i\gamma\ .
\end{equation}
Now, we are going to find the spectrum together with the
corresponding eigenfunctions:
\begin{equation}\label{}
  H_{\rm dho} \star F = F \star H_{\rm dho} = EF\ .
\end{equation}
Define $F^\pm_{00}$  to be functions corresponding to   ``$\pm$
ground states'', that is,
\begin{equation}\label{a1a2*}
a_1 \star F^+_{00}=0\ , \ \ \ \ F^+_{00} \star a^*_2 =0 \ ,
\end{equation}
and
\begin{equation}\label{a2a1*}
a_2 \star F^-_{00}=0\ , \ \ \ \ F^-_{00} \star a^*_1 =0 \ .
\end{equation}
 Unique normalized solutions of (\ref{a1a2*})--(\ref{a2a1*}) are given by:
\begin{equation}\label{}
  F^+_{00} = \frac{1}{(2\pi \hbar)^2}\,
  e^{\frac{2i}{\hbar}(x_1p_1+x_2p_2)}\ ,
\end{equation}
and
\begin{equation}\label{}
  F^-_{00} = \overline{F^+_{00}}\ .
\end{equation}
 Moreover, defining
\begin{equation}\label{}
  F^+_{nm} \ \propto\ (a_2^*)^n \star F^+_{00} \star a_1^m\ ,
\end{equation}
and
\begin{equation}\label{}
  F^-_{nm} \ \propto\ (a_1^*)^n \star F^-_{00} \star a_2^m\ ,
\end{equation}
 one shows
\begin{equation}\label{}
 H_{\rm dho} \star F^+_{nm} = F^+_{nm} \star H_{\rm dho} = E_{nm}F^+_{nm}\ ,
\end{equation}
and
\begin{equation}\label{}
 H_{\rm dho} \star F^-_{nm} = F^-_{nm} \star H_{\rm dho} = \overline{E_{nm}} F^-_{nm}\ ,
\end{equation}
 with
\begin{equation}\label{Enm}
 E_{nm} = \hbar\alpha\left(m+ \frac 12 \right) -
 \hbar\overline{\alpha}\left( n + \frac 12 \right)
 = \hbar \omega (m-n) - i\hbar\gamma(n+m+1)\ .
\end{equation}
Let us compare the above formulation with the standard operator
approach (see \cite{Kossak1}, \cite{Koopman}) based on the
following Hamilton operator:
\begin{equation}\label{H-osc-d-q}
  \widehat{H}_{\rm dho} = \omega( \widehat{p}_1\widehat{x}_2 - \widehat{p}_2\widehat{x}_1 ) -
  \frac{\gamma}{2} ( \widehat{p}_1\widehat{x}_1 + \widehat{x}_1\widehat{p}_1 +
  \widehat{p}_2\widehat{x}_2 + \widehat{x}_2\widehat{p}_2 ) \ .
\end{equation}
Following  (\ref{a1})--(\ref{a2}) we introduce
$(\widehat{a}_k,\widehat{a}^*_k)$ which satisfy (\ref{CCR1}) and
(\ref{CCR2}) with Moyal bracket $\{\ , \ \}_{\rm M}$ replaced by
the commutator. Now, let us introduce  ``$\pm$ ground states"
$\varphi^\pm_{00}$ as the states satisfying:
\begin{equation}\label{a1a1*}
  \widehat{a}_1 \varphi^+_{00} = \widehat{a}^*_1 \varphi^+_{00} = 0 \ ,
\end{equation}
 and
\begin{equation}\label{a2a2*}
  \widehat{a}_2 \varphi^-_{00} = \widehat{a}^*_2 \varphi^-_{00} = 0 \ .
\end{equation}
Moreover, define two families of excited states:
\begin{equation}\label{}
  \varphi^+_{nm} := \widehat{a}_2^n (\widehat{a}_2^*)^m \, \varphi^+_{00} \
  ,
\end{equation}
and
\begin{equation}\label{}
  \varphi^-_{nm} := \widehat{a}_1^n (\widehat{a}_1^*)^m \, \varphi^-_{00} \
  .
\end{equation}
It is easy to show that
\begin{equation}\label{}
  \widehat{H}_{\rm dho} \varphi^+_{nm} = E_{nm} \varphi^+_{nm} \ ,
\end{equation}
and
\begin{equation}\label{}
  \widehat{H}_{\rm dho} \varphi^-_{nm} = \overline{E_{nm}} \varphi^-_{nm} \ ,
\end{equation}
with $E_{nm}$ defined in (\ref{Enm}).  Using standard
$(x_1,x_2)$-representation, i.e. $\widehat{x}_k \varphi =
x_k\varphi$ and $\widehat{p}_k = -i\hbar\partial_k\varphi$, one
easily solves (\ref{a1a1*}) and (\ref{a2a2*}). Up to non-important
constants one obtains:
\begin{equation}\label{00+}
  \varphi^+_{00}(x_1,x_2) = \delta(x_1)\delta(x_2) \ ,
\end{equation}
and
\begin{equation}\label{00-}
  \varphi^-_{00}(x_1,x_2) = 1\ .
\end{equation}
Clearly, neither $\varphi^+_{00}$ nor $\varphi^-_{00}$ belong to
$L^2( \mathbb{R}^2)$. Note, that there is a striking similarity
between $F^\pm_{nm}$, $\varphi^\pm_{nm}$ and $F^\pm_n$,
$\varphi^\pm_n$ from section~\ref{TOY} Now, using a pair of
resonant states $\varphi^\pm_{nm}$ one may easily show that
\begin{equation}\label{}
  F^\pm_{nm} \ \propto\ \int d{\bf y} \, e^{-i{\bf p}{\bf y}}
  \overline{ \varphi^\mp_{nm}}({\bf x} - \frac{\hbar}{2}{\bf y})
  \varphi^\pm_{nm}({\bf x} + \frac{\hbar}{2}{\bf y}) \ .
\end{equation}
Moreover, using a straightforward algebra one may prove the
following

\noindent {\bf Proposition.} {\em Normalized functions
$F^\pm_{nm}$ satisfy:}
\begin{eqnarray}
\pi^\pm_{{\rm x}|nm}({\bf x}) &=& \int F^\pm({\bf x},{\bf p})\,
d{\bf p} =  \delta({\bf x})\ ,
\\ \pi^\pm_{{\rm p}|nm}({\bf p}) &=& \int F^\pm({\bf x},{\bf p})\,
d{\bf x} = \delta({\bf p})\ ,
\end{eqnarray}
{\em and hence}
\begin{equation}\label{}
  \int F^\pm_{nm}({\bf x},{\bf p})\,  d{\bf x}\,d{\bf p} = 1\ .
\end{equation}
{\em Moreover, }
\begin{equation}\label{}
  F^\pm_{nm} \star F^\pm_{kl} = \frac{1}{(2\pi \hbar)^2}\,
  \delta_{nk}\delta_{ml}\, F^\pm_{nm}\ ,
\end{equation}
{\em in analogy to (\ref{WnWm}). This  implies the following
resolution of identity:}
\begin{equation}\label{}
  \sum_{n,m} F^\pm_{nm} =  \frac{1}{(2\pi \hbar)^2}\ ,
\end{equation}
{\em in accordance to (\ref{res}).}

\section{Another representation}
\setcounter{equation}{0}

Both oscillator Wigner functions $W_n$ and the corresponding
$F^\pm_n$ and $F^\pm_{nm}$ from sections \ref{TOY} and
\ref{DAMPED-HO} respectively,  are stationary function, i.e. they
commute with the corresponding Hamiltonian. In the case of $W_n$
and $F^\pm_n$ this property follows from that fact that
\begin{equation}\label{}
  W_n=W_n(H_{\rm ho}) \ \ \ \ \ {\rm and} \ \ \ \ \
  F^\pm_n=F^\pm(H_{\rm d})\ .
\end{equation}
Now, in the case of a damped harmonic oscillator the corresponding
Hamiltonian (\ref{H-osc-d}) may be written as a sum
\begin{equation}\label{}
  H_{\rm dho} = H_1 + H_2\ ,
\end{equation}
where
\begin{equation}\label{}
  H_1 = \omega\, {\bf p} \wedge {\bf x}\ \ \ \ \ {\rm and} \ \ \ \
  \ H_2 = - \gamma {\bf p} \cdot {\bf x} \ .
\end{equation}
Clearly,
\begin{equation}\label{}
  F^\pm_{nm} = F^\pm_{nm}(H_2) \ ,
\end{equation}
and the stationarity of $F^\pm_{nm}$ follows from
\begin{equation}\label{}
  \{H_1,H_2\}_{\rm M}=0\ .
\end{equation}
Now, we show that  it is possible to construct another family
$G_{nm}$ such that
\begin{equation}\label{}
  G_{nm}=G_{nm}(H_1)\ ,
\end{equation}
and $G_{nm}$ satisfy the corresponding eigen-problem
\begin{equation}\label{}
  H_{\rm dho} \star G_{nm} = G_{nm} \star H_{\rm dho} =
  \mu_{nm} G_{nm}\  .
\end{equation}
Let us define $G_{00}$ as a ``ground state" satisfying
\begin{equation}\label{a-G00}
  a_k \star G_{00}=0\ \ \ \ {\rm and} \ \ \ \ G_{00} \star a^*_k =0 \ , \ \ \ \
  k=1,2\ .
\end{equation}
Solving (\ref{a-G00}) one finds:
\begin{equation}\label{G00}
  G_{00} = e^{\frac{2}{\hbar}(x_1p_2 - x_2p_1)} =
  e^{\frac{2}{\hbar}\, {\bf x} \wedge {\bf p}} \ .
\end{equation}
Clearly, $G_{00}$, contrary to $F^\pm_{00}$,  is not integrable
over $ \mathbb{R}^4$. Using (\ref{CCR1}) and (\ref{CCR2}) it is
easy to show that the following set of functions:
\begin{equation}\label{Gnm}
  G_{nm} = (a^*_1)^n \star (a^*_2)^n \star G_{00} \star
  a_2^n \star a_1^m \ ,
\end{equation}
satisfy
\begin{equation}\label{}
 H_{\rm dho} \star G_{nm} = G_{nm} \star H_{\rm dho} = \mu_{nm}G_{nm}\ ,
\end{equation}
with
\begin{equation}\label{mu-nm}
 \mu_{nm} = \hbar\alpha\left(n+ \frac 12 \right) +
 \hbar\overline{\alpha}\left( m + \frac 12 \right)
 = \hbar \omega (n+m+1) - i\hbar\gamma(n-m)\ .
\end{equation}
 Note,
that
\begin{equation}\label{}
  G_{nm} = \overline{G_{mn}}\ ,
\end{equation}
and
\begin{equation}\label{}
 \mu_{nm} = \overline{\mu_{mn}}\ .
\end{equation}
Therefore, we have a natural pairing $(G_{nm},G_{mn})$ in analogy
to $(F^+_{nm},F^-_{nm})$. Interestingly, both approaches give
completely different spectra of $\widehat{H}_{\rm dho}$: $E_{nm}$
and $\mu_{nm}$ defined in (\ref{Enm}) and (\ref{mu-nm}),
respectively.

Let us compare the above formulation with the standard operator
approach (see \cite{Kossak1}, \cite{Koopman}) based on the
Hamilton operator (\ref{H-osc-d-q}).
 The commutation relations may be easily
represented in the space of functions of two variables
$(x_1,p_2)$:
\begin{eqnarray}\label{}
  \widehat{a}_1 &=& \frac{1}{\sqrt{2\hbar}} \left( x_1 - \hbar
  \frac{\partial}{\partial p_2} \right) \ , \hspace{1.3cm}
\widehat{a}^*_1 \,=\, \frac{1}{\sqrt{2\hbar}} \left( x_1 + \hbar
  \frac{\partial}{\partial p_2} \right) \ , \\
\widehat{a}_2 &=& \frac{1}{\sqrt{2\hbar}} \left( -p_2 + \hbar
  \frac{\partial}{\partial x_1} \right) \ , \hspace{1cm}
\widehat{a}^*_2 \,=\, \frac{1}{\sqrt{2\hbar}} \left( -p_2 - \hbar
  \frac{\partial}{\partial x_1} \right) \ .
\end{eqnarray}
Introducing a ground state $\psi_{00}$:
\begin{equation}\label{}
  \widehat{a}_1\psi_{00} =  \widehat{a}_2\psi_{00}=0\ ,
\end{equation}
one finds:
\begin{equation}\label{}
  \psi_{00}(x_1,p_2) = e^{x_1p_2/\hbar} \ .
\end{equation}
Defining
\begin{equation}\label{}
  \psi_{nm} = (\widehat{a}^*_1)^n(\widehat{a}^*_2)^m\psi_{00} \ ,
\end{equation}
one shows
\begin{equation}\label{}
  \widehat{H}_{\rm dho}\psi_{nm} = \mu_{nm}\psi_{nm} \ ,
\end{equation}
with $\mu_{nm}$ given by (\ref{mu-nm}). One may show that it is
possible to construct $G_{nm}$ defined in (\ref{Gnm}) out of
resonant states $\psi_{nm}$. However, contrary to $F^\pm_{nm}$,
$G_{nm}$ are not normalizable and the striking analogy with Wigner
functions is lost.

\section{Concluding remarks}

In the present paper we analyzed the quantization of simple
classical damped systems: a toy model defined by $\dot{x} =
-\gamma x$ and the damped harmonic oscillator. Both systems give
rise to resonant states and the corresponding energy spectra are
discrete and complex.  It turns out that resonant states appear
always in pairs: if $\psi_1$ corresponds to $E$ then there exists
$\psi_2$ corresponding to $\overline{E}$. We showed that each pair
of such states may be used to construct an analog of the
stationary Wigner function. Actually one constructs a pair of
stationary functions
\begin{equation}\label{}
  F \propto \int dy e^{-ipy} \overline{\psi}_1(x -
  \frac{\hbar}{2}y) \, \psi_2(x + \frac{\hbar}{2}y)\ \ \  \ \ \
  {\rm and}\ \ \  \ \ \ \overline{F}\ . \nonumber
\end{equation}
 A slightly different approach to
quantization of damped oscillator was applied in \cite{Vitiello}.
In the forthcoming paper we show that both approaches are closely
related. In a different context a quantum damped harmonic
oscillator was recently analyzed in \cite{Benatti}.

\section*{Acknowledgements}

It is pleasure to thank professor Andrzej Kossakowski for very
interesting  discussions.  This work was partially supported by
the Polish State Committee for Scientific Research (KBN) Grant no
2P03B01619.

\end{document}